\documentclass[preprint]{imsart}
\makeatletter
\let\c@author\relax
\makeatother
\usepackage{biblatex}
\RequirePackage[OT1]{fontenc}
\RequirePackage{amsthm,amsmath}
\RequirePackage[colorlinks,citecolor=blue,urlcolor=blue]{hyperref}

\usepackage{graphicx}
\usepackage[T1]{fontenc}
\usepackage[utf8]{inputenc}
\usepackage{tikz}
\usepackage{bm}
\usepackage{pgfplots}
\usepackage{lastpage}
\usepackage{wrapfig}
\usetikzlibrary{fit,positioning}
\usepackage{geometry}
\usepackage{dsfont}
\usepackage{amssymb,relsize}
\usepackage{kbordermatrix}
\usepackage{mathtools}

\makeatletter
\newcommand{\subalign}[1]{%
  \vcenter{%
    \Let@ \restore@math@cr \default@tag
    \baselineskip\fontdimen10 \scriptfont\tw@
    \advance\baselineskip\fontdimen12 \scriptfont\tw@
    \lineskip\thr@@\fontdimen8 \scriptfont\thr@@
    \lineskiplimit\lineskip
    \ialign{\hfil$\m@th\scriptstyle##$&$\m@th\scriptstyle{}##$\crcr
      #1\crcr
    }%
  }
}
\makeatother

\newcommand{\etal}{\textit{et al}.}

\DeclareMathOperator*{\cartes}{\mathlarger{\mathlarger{\mathlarger{\mathlarger{\mathbf{\otimes}}}}}}

\newcommand{\dens}{} 
\newcommand{\E}{\bm{E}}

\newcommand{\mcmc}{\zeta}
\newcommand{\Mcmc}{\bm{\mcmc}}
\newcommand{\mcmi}{\iota}
\newcommand{\Mcmi}{\bm{\mcmi}}
\newcommand{\hist}{\eta}
\newcommand{\Hist}{\bm{\hist}}

\newcommand{\ball}{\mathcal{B}}
\newcommand{\prog}{\mathsf{P}}
\newcommand{\samp}{\mathsf{S}}
\newcommand{\cond}{\mathsf{C}}
\newcommand{\augm}{\mathsf{A}}
\newcommand{\fink}{\mathsf{R}}

\newcommand{\xtt}{x}
\newcommand{\Xtt}{\bm{\xtt}}
\newcommand{\xt}{z}
\newcommand{\Xt}{\bm{\xt}}
\newcommand{\yt}{y}
\newcommand{\Yt}{\bm{\yt}}
\newcommand{\raw}{\tilde{z}}
\newcommand{\Raw}{\bm{\raw}}
\newcommand{\norm}{|||}

\newcommand{\subdim}{K}
\newcommand{\sell}{{\{l\}}} 

\newcommand{\bla}{{\alpha}} 
\newcommand{\blb}{{\beta}} 
\newcommand{\kfa}{{a}} 
\newcommand{\kfb}{{b}} 
\newcommand{\kfc}{{c}} 
\newcommand{\kfd}{{q}} 
\newcommand{\kfe}{{e}} 

\newcommand{\weight}{{v}}

\startlocaldefs
\numberwithin{equation}{section}
\theoremstyle{plain}

\endlocaldefs

\begin{document}

\begin{frontmatter}
\title{A High-Dimensional Particle Filter Algorithm}
\runtitle{High-d Particle Filter}

\begin{aug}
\author{\fnms{Jameson} \snm{Quinn}\thanksref{t1}\ead[label=e1]{jameson.quinn@gmail.com}},

\thankstext{t1}{Thanks to Mira Bernstein, Pierre Jacob, and Luke Miratrix for constructive criticism of this manuscript.}
\runauthor{J. Quinn}

\affiliation{Harvard University\thanksmark{m1}}


\end{aug}

\begin{abstract}

Online data assimilation in time series models over a large spatial extent is an important problem in both geosciences and robotics. Such models are intrinsically high-dimensional, rendering traditional particle filter algorithms ineffective. Though methods that begin to address this problem exist, they either rely on additional assumptions or lead to error that is spatially inhomogeneous. I present a novel particle-based algorithm for online approximation of the filtering problem on such models, using the fact that each locus affects only nearby loci at the next time step. The algorithm is based on a Metropolis-Hastings-like MCMC for creating hybrid particles at each step. I show simulation results that suggest the error of this algorithm is uniform in both space and time, with a lower bias, though higher variance, as compared to a previously-proposed algorithm.
\end{abstract}



\end{frontmatter}

\section{Background}

Filtering problems arise in many applied contexts, whenever noisy observations over time must be combined, using an explicit dynamical model, into a best-guess distribution of a current state. In cases where the system being modeled involves processes over a large spatial extent, such as models of weather or other large-scale fluid dynamics, filtering is also called data assimilation.\cite{apteDataAssimilationMathematical2008} 
This is an active area of research, with broad applications in predictive geoscience\cite{bengtssonCurseofdimensionalityRevisitedCollapse2008}\cite{vanleeuwenNonlinearDataAssimilation2015} and robotics\cite{stachnissSimultaneousLocalizationMapping2016}. In fact, it is considered to be among the central problems in both of these disciplines.\footnote{
According to the papers cited above, there is ``much focus in the [geoscience] literature on the assimilation of data and numerical models pertain[ing] to the sampling of high-dimensional probability density functions'' \cite{bengtssonCurseofdimensionalityRevisitedCollapse2008}, and ``The SLAM [simultaneous location and mapping] problem is generally regarded as one of the most important problems in the pursuit of building truly autonomous mobile robots.''\cite{stachnissSimultaneousLocalizationMapping2016}} 

\begin{figure}
\begin{tikzpicture}
\tikzstyle{main}=[circle, minimum size = 5mm, thick, draw =black!80, node distance = 8mm]
\tikzstyle{connect}=[-latex, thick]
\tikzstyle{box}=[rectangle, draw=black!270]
  \node[main, fill = white!100] (x0) [label=above:$X_0$] { };
  \node[main] (x1) [right=of x0,label=above:$X_1$] { };
  \node[main] (x2) [right=of x1,label=above:$\dots$] {};
  \node[main] (x3) [right=of x2,label=above:$X_{T-1}$] { };
  \node[main] (x4) [right=of x3,label=above:$X_T$] { };
  \node[main, fill = black!27] (y0) [below=of x0,label=below:$Y_0$] { };
  \node[main, fill = black!27] (y1) [below=of x1,label=below:$Y_1$] { };
  \node[main, fill = black!27] (y2) [below=of x2,label=below:$\dots$] { };
  \node[main, fill = black!27] (y3) [below=of x3,label=below:$Y_{T-1}$] { };
  \node[main, fill = black!27] (y4) [below=of x4,label=below:$Y_T$] { };
  \path (x0) edge [connect] (x1)
        (x1) edge [connect] (x2)
		(x2) edge [connect] (x3)
		(x3) edge [connect] (x4)
		(x0) edge [connect] (y0)
		(x1) edge [connect] (y1)
		(x2) edge [connect] (y2)
		(x3) edge [connect] (y3)
		(x4) edge [connect] (y4)
		;
\end{tikzpicture}
\caption[]{Graphical model of a low-dimensional filtering problem.}
\label{lowdim}
\end{figure}
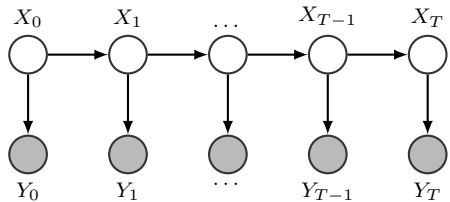

The basic filtering problem is as follows. We model the state of the system at time $t$ as a random variable $X_t$. In our context, $X_t$ will have values $x_{\subalign{l\\t}}$ at each spacial locus $l$, for a large number of loci. We assume $X_0,...,X_T$ form a Markov chain with known and sampleable densities for both the initial state ($\pi_0$) and transition function ($\prog$, which maps states or densities at time $t-1$ to densities at time $t$). We also have a series of observations $Y_1,...,Y_T$, and we assume that each $Y_t$ depends only on the corresponding $X_t$ according to a known and sampleable observation density $f(Y_t|X_t)$. This is shown graphically in Figure~\ref{lowdim}.

At each time step $t$, we wish to estimate the filtering distribution: that is, the probability density $\pi_t$ of $X_t|\{Y_1,...,Y_t\}$. Because the $X_t$ are Markovian, and $Y_t$ depends only $X_t$, we can write $\pi_t$ recursively as
\begin{equation}
\pi_t(\cdot)= \E_{X_{t-1}\sim\pi_{t-1}}[P(X_t\in\cdot|Y_t,X_{t-1})]
\end{equation}

To minimize subscripts, we adopt the following notation:
\begin{itemize}
    \item We abbreviate the filtering distributions $\pi_{t-1}$ and $\pi_t$ by $\tau$ and $\pi$ respectively.
    \item We abbreviate samples from $X_{t-1}$ and $X_t$ by $\Xtt$ and $\Xt$ respectively.
    \item When $\bm{y}_{t-1}$ is not relevant, we write $\Yt$ for $\bm{y}_t$.
    \item Superscripts should not be read as exponentiation for these and similar entities.
\end{itemize}  

The recursive formula for $\pi_t$ suggests the possibility of online calculation, with only a constant computing time required to update from $\tau = \pi_{t-1}$ to $\pi = \pi_t$. However, unless we assume a particular parametric form for $\pi$, there is no finite set of sufficient statistics that could stand in for the full distribution. Thus, aside from very simple special cases, exact calculation is impossible; we look for an approximation instead.

A widely-used recursive algorithm for approximating the filtering distribution is the bootstrap particle filter. Assuming we have a sampleable distribution $\hat\tau$ at time $t-1$ that approximates the true filtering distribution $\tau$, we proceed as follows:
\begin{enumerate}
\item Sample $M$ iid particles $\Xtt^{1..M}$ from $\hat\tau$. (Note that if we have been following the algorithm up to step $t-1$, then $\hat\tau$ takes the form of step (3) below.)
\item For each $\Xtt^{i}$, progress it to get candidate particle $\Xt^{i}\sim \prog \Xtt^i$.
\item Find weights $w^{i}\equiv f(\Yt|\Xt^{i})$. The set of weighted particles forms 
$$\hat\pi\equiv\frac{\sum_{i=1}^M w^{i}\delta(\Xt^i)}{\sum_{i=1}^M w^{i}}.$$ 
Here $\delta(a)$ is the Dirac delta density; for example, $\frac12(\delta(0)+\delta(1))$ is the Bernoulli distribution with $p=\frac12$. 
\end{enumerate} 

A key property of the particle filter algorithm is that, for large enough $M$, the Monte Carlo error remains under control, even as the time steps accumulate.\cite{cappeInferenceHiddenMarkov2009} Specifically, let $\hat\pi_t^M$ be the approximation to $\pi_t$ obtained using $M$ particles, as above. Let $\mathcal{F}$ be the set of functions from the domain of $\pi_t$ to $(-1,1)$. For each $\mathrm{f} \in \mathcal{F}$, denote $\E_{X\sim{\pi_t}}(\mathrm{f}(X))$ and $\E_{X\sim{\hat\pi_t^M}}(\mathrm{f}(X))$ by $\pi_t(\mathrm{f})$ and $\hat\pi_t^M(\mathrm{f})$ respectively. Then 
\begin{equation} \label{basicbounds}
\sup_\mathcal{F} \E|\pi_t(\mathrm{f})-\hat\pi^M_t(\mathrm{f})|\leq\frac{C}{\sqrt{M}},
\end{equation}
for some constant $C$ that {\em does not depend on $t$}.
The outer expectation here is taken over the randomness of the algorithm itself; that is, considering the distribution $\hat\pi$ as itself a random variable, while $\pi$ is fixed.\cite[p.~2814]{rebeschiniCanLocalParticle2015}

Now suppose that we are interested in modeling processes with a large spatial extent. For instance, in a weather model, one might use a lattice of points that cover the region of interest, with various continuous values (temperature, humidity, pressure, wind, etc.) recorded at each locus. If $X_t$ contains information about $L$ separate spatial loci, and
the state space at each locus has dimension $\subdim$, then the full state space of $X_t$ has dimension $\subdim L$. In practical applications, this can easily be $10^7$ or more.\cite{vanleeuwenParticleFilteringGeophysical2009} 

\begin{figure}
\begin{tikzpicture}
\tikzstyle{main}=[circle, minimum size = 8mm, thick, draw =black!80, node distance = 12mm]
\tikzstyle{connect}=[-latex, thick]
\tikzstyle{box}=[rectangle, draw=black!270]
  \node[main, fill = white!100] (x0) [] {$\scriptstyle{ x}_{\substack{\scriptstyle{0}\\\scriptstyle{0}}}$ };
  \node[main] (x1) [right=of x0,label=above:] { $\scriptstyle{ x}_{\substack{\scriptstyle{0}\\\scriptstyle{1}}}$ };
  \node[main] (x2) [right=of x1,label=above:] {$\dots$};
  \node[main] (x3) [right=of x2,label=above:] {$\scriptstyle{ x}_{\substack{\scriptstyle{0}\\\scriptstyle{T}}}$ };
  
  \node[main] (x01) [above right=1mm of x0,label=above:] { $\scriptstyle{ x}_{\substack{\scriptstyle{1}\\\scriptstyle{0}}}$ };
  \node[main] (x11) [above right=1mm of x1,label=above:] {$\scriptstyle{ x}_{\substack{\scriptstyle{1}\\\scriptstyle{1}}}$ };
  \node[main] (x21) [above right=1mm of x2,label=above:] {$\dots$ };
  \node[main] (x31) [above right=1mm of x3,label=above:] {$\scriptstyle{ x}_{\substack{\scriptstyle{1}\\\scriptstyle{T}}}$ };
  
  \node[main] (x02) [above right=1mm of x01,label=above:] {$\dots$ };
  \node[main] (x12) [above right=1mm of x11,label=above:] {$\dots$  };
  \node[main] (x22) [above right=1mm of x21,label=above:] {$\dots$ };
  \node[main] (x32) [above right=1mm of x31,label=above:] {$\dots$  };
  
  \node[main] (x03) [above right=1mm of x02,label=above:] { $\scriptstyle{ x}_{\substack{\scriptstyle{L}\\\scriptstyle{0}}}$};
  \node[main] (x13) [above right=1mm of x12,label=above:] {$\scriptstyle{ x}_{\substack{\scriptstyle{L}\\\scriptstyle{1}}}$ };
  \node[main] (x23) [above right=1mm of x22,label=above:] {$\dots$ };
  \node[main] (x33) [above right=1mm of x32,label=above:] {$\scriptstyle{ x}_{\substack{\scriptstyle{L}\\\scriptstyle{T}}}$ };
  
  \node[main, fill = black!27] (y0) [below=of x0,label=below:]   { $\scriptstyle{ y}_{\substack{\scriptstyle{0}\\\scriptstyle{0}}}$ };
  \node[main, fill = black!27] (y1) [below=of x1,label=below:]   { $\scriptstyle{ y}_{\substack{\scriptstyle{0}\\\scriptstyle{1}}}$ };
  \node[main, fill = black!27] (y2) [below=of x2,label=below:]   { $\dots$ };
  \node[main, fill = black!27] (y3) [below=of x3,label=below:]   { $\scriptstyle{ y}_{\substack{\scriptstyle{0}\\\scriptstyle{T}}}$ };

  \node[main, fill = black!27] (y01) [below=of x01,label=below:]   { $\scriptstyle{ y}_{\substack{\scriptstyle{1}\\\scriptstyle{0}}}$ };
  \node[main, fill = black!27] (y11) [below=of x11,label=below:]   { $\scriptstyle{ y}_{\substack{\scriptstyle{1}\\\scriptstyle{1}}}$ };
  \node[main, fill = black!27] (y21) [below=of x21,label=below:]   { $\dots$ };
  \node[main, fill = black!27] (y31) [below=of x31,label=below:]   { $\scriptstyle{ x}_{\substack{\scriptstyle{1}\\\scriptstyle{T}}}$ };

  \node[main, fill = black!27] (y02) [below=of x02,label=below:]   { $\dots$  };
  \node[main, fill = black!27] (y12) [below=of x12,label=below:]   { $\dots$  };
  \node[main, fill = black!27] (y22) [below=of x22,label=below:]   { $\dots$ };
  \node[main, fill = black!27] (y32) [below=of x32,label=below:]   { $\dots$  };

  \node[main, fill = black!27] (y33) [below=of x33,label=below:]   {$\scriptstyle{ y}_{\substack{\scriptstyle{L}\\\scriptstyle{t}}}$  };
  \path 
    (x0) edge [connect] (x1)
    (x1) edge [connect] (x2)
		(x2) edge [connect] (x3)
		
    (x0) edge [connect] (x11)
    (x1) edge [connect] (x21)
		(x2) edge [connect] (x31)

    (x01) edge [connect] (x1)
    (x11) edge [connect] (x2)
		(x21) edge [connect] (x3)
		
    (x01) edge [connect] (x11)
    (x11) edge [connect] (x21)
		(x21) edge [connect] (x31)
		
    (x01) edge [connect] (x12)
    (x11) edge [connect] (x22)
		(x21) edge [connect] (x32)

    (x02) edge [connect] (x11)
    (x12) edge [connect] (x21)
		(x22) edge [connect] (x31)
		
    (x02) edge [connect] (x12)
    (x12) edge [connect] (x22)
		(x22) edge [connect] (x32)
		
    (x02) edge [connect] (x13)
    (x12) edge [connect] (x23)
		(x22) edge [connect] (x33)

    (x03) edge [connect] (x12)
    (x13) edge [connect] (x22)
		(x23) edge [connect] (x32)
		
    (x03) edge [connect] (x13)
    (x13) edge [connect] (x23)
		(x23) edge [connect] (x33)

		(x0) edge [connect] (y0)
		(x1) edge [connect] (y1)
		(x2) edge [connect] (y2)
		(x3) edge [connect] (y3)
		
		(x01) edge [connect] (y01)
		(x11) edge [connect] (y11)
		(x21) edge [connect] (y21)
		(x31) edge [connect] (y31)

		(x02) edge [connect] (y02)
		(x12) edge [connect] (y12)
		(x22) edge [connect] (y22)
		(x32) edge [connect] (y32)

		(x33) edge [connect] (y33)
		;
\end{tikzpicture}
\caption[]{Graphical model of a (simple) high-dimensional filtering problem.}
\label{highdim}
\end{figure}
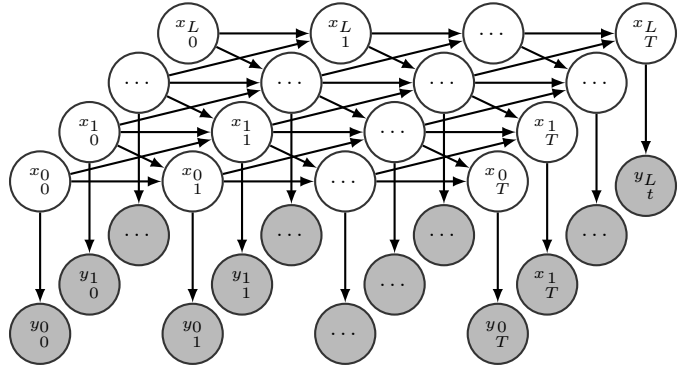

To see why, consider a schematic diagram of the model (Figure~\ref{highdim}), where the state of the system at time $t$ and locus $l$ is denoted $x_{\subalign{l\\t}}$. 

Note the following assumptions implicit in the diagram: \begin{itemize}
    \item $(y_{\substack{l\\t}}|x_{\substack{l\\t}})\perp\!\!\!\perp x_{\substack{k\neq l\\t}}$. This assumption will be used in the following for simplicity, although I believe it can be relaxed in practice with only minor additional complications. 
    \item More crucially, $x_{\substack{l\\t}}$ depends on $x_{\substack{k\\t-1}}$ only for $k$ in some small spatial ``neighborhood" $\mathcal{N}(l)$ of $l$. The precise composition of $\mathcal{N}(l)$ depends on model assumptions as well as the way that the $L$ loci are positioned in space. The diagram depicts the case where the loci are all laid out along a single spatial dimension, so that the immediate neighbors of locus $l$ are loci $l-1$ and $l+1$. In practical applications, the loci would more likely be connected in a 2D or 3D grid.
\end{itemize}
I will discuss this locality of dynamics assumption further below, as it is key to the performance of the algorithm I propose in this paper.

Technically speaking, the error bounds in \ref{basicbounds} still apply: errors are stable over time and inversely proportional to the square root of the number of particles ($\propto 1/\sqrt{M}$). But it is widely recognized that the bootstrap particle filter is no longer a practical solution in this context, due to weight degeneracy.\cite{snyderObstaclesHighDimensionalParticle2008}\cite{bickelSharpFailureRates2008}\cite{agapiouImportanceSamplingIntrinsic2017} The problem is that, in the resampling step, the majority of the weight will tend to be carried by only a small fraction of the proposed particles. To see why, note that the log likelihood of each particle is the sum of its log likelihood at each locus. Since these terms are roughly independent, as $L$ increases, the empirical distribution of log likelihoods of the particles comes to resemble a Gaussian distribution with variance that scales linearly with $L$. Thus the weights come to be distributed approximately according to a log-normal distribution, whose skewness increases exponentially with $L$. Thus, the fraction of particles with above-average weight will shrink exponentially with $L$. \footnote{Bickel \etal\cite{bickelSharpFailureRates2008} formalize this line of argument, showing that the variance of the log likelihood may be seen as an estimate of the effective state dimension depicted by the measurements $\Yt$.}

In fact, without an exponentially large number of particles, not only will one of them tend to have more weight than all the others, but it is likely that there is some missing value whose weight would dominate even our best particle. At this point the particle filter ceases to be a useful approximation of the filtering distribution. That is, although the constant $C$ in Equation~\ref{basicbounds} does not depend on time, it does grow exponentially with $L$. This is what is known as {\em curse of dimensionality} for particle filters. 

\subsection{Existing state of the art (Rebeschini and van Handel, 2015)}

There have been various proposals for dealing with this curse of dimensionality in general. In fact, there are three different recent survey articles reviewing and comparing these, by Septier and Peters\cite{septierOverviewRecentAdvances2015}, Morzfield \etal\cite{morzfeldLocalizationMCMCSampling2017}, and Farchi and Bocquet.\cite{farchiReviewArticleComparison2018} Some of these prior methods do not use the locality of dynamics assumption, which I believe limits their effectiveness. This includes Gilks and Berzuini,\cite{gilksFollowingMovingTarget2001} who suggest rejuvenating particles with MCMC steps, targeted to the filtering distribution, to avoid the duplication problem from resampling; and Goodsill and Clapp,\cite{godsillImprovementStrategiesMonte2000} who propose using bridging densities such as annealed quasi-filtering densities to solve the problem of lack of overlap between the progressed density $\prog\tau\equiv(\Xt|\Yt_1,...,\Yt_{t-1})$ and the likelihood $f(\Yt_t|\Xt)$. 

The two previous proposals that do use locality of dynamics come from  Poterjoy\cite{poterjoyLocalizedParticleFilter2015}\cite{poterjoyConvectiveScaleDataAssimilation2017} and from Rebeschini and van Handel.\cite{rebeschiniCanLocalParticle2015} Of these two, Rebeschini and van Handel's proposal is more generally applicable, so I will explain it further below. Poterjoy, on the other hand, suggests a scheme that, as given, is limited to situations of sparse observations; it uses estimated covariance matrices to blend resampled particle filter values at a local scale with unresampled values at a meso-scale.

Rebeschini and van Handel's proposal is called the block particle filter; also sometimes termed the localized particle filter. In simple terms, they replace the global resampling step of the bootstrap particle filter with a local resampling step which constructs new particles by resampling neighborhoods independently.

They begin their theoretical discussion by offering an overall point of view of the problem which very much inspired the current work. They focus on the decay of correlations between local values as spatial distance increases, which they say is ``in essence a spatial counterpart of the much better-understood [temporal] stability property of nonlinear filters''. This decay of correlations, discussed further below, is a product of the locality-of-dynamics assumption shown in the diagrams above. 

Rebeschini and van Handel partition the loci ${1,...,L}$ of the progressed particles into $J$ zones $\{\mathcal{Z}_j\}$, where each zone consists of a small number of (contiguous) loci. They then weight and resample values from each zone independently. This produces what I would call ``Frankenstein'' particles, sewn together from pieces which tend to fit well with observations locally, but which may often come from progressing different particles from time $t-1$.

The precise steps of the block particle filter algorithm are as follows:

\begin{enumerate}
\item  Given $\hat\tau$, a sampleable distribution that approximates the ideal filtering distribution $\tau$,
  sample $M$ iid particles $\Xtt^{1..M}$ from $\hat\tau$. (Note that if $\hat{\tau}$ takes the form of the output of step (3) below,
  then this amounts to sampling, independently with replacement, a $k^i_j$ for
  each zone $Z_j;1\leq j\leq J$ and particle $i$ with probability $w^{k^i_j}_{Z_j}/\Sigma_n w^{n}_{Z_j}$; then putting those together to make the final particles, so that $l\in Z_j\Rightarrow \xt^i_l=\xt^{k^i_j}_l$.)
\item For each $\Xtt^i$, progress it to get $\Xt^{i}\sim \prog \Xtt$.
\item Find weights for each particle for each zone: $w^{i}_{Z_j}={\prod_{l\in Z_j}}f(\dens y_l|\xt^{i}_l)$. Then 
$$\hat\pi\equiv\cartes_{j=1}^{J} \frac{\sum_{i=1}^{M} w^{i}_{Z_j}\delta(\Xtt^i_{Z_j})}{\sum_{i=1}^{M} w^{i}_{Z_j}}$$
\end{enumerate}

Intuitively, cutting the progressed particles into zones and reconstituting them is a way to solve the exponential curse of dimensions. However, it gives rise to a different problem: it breaks any covariances across zone boundaries, whether those come from covariances at the $t-1$ time step or are induced by the transition kernel using a history that overlaps the boundary. These broken covariances lead to error at the boundaries,
which does not disappear even as number of particles goes to infinity.

It could also then lead to unrealistic dynamics near the boundaries at later time steps, especially if the forward density operator $\prog$ is nonlinear. \cite[p.~2829]{rebeschiniCanLocalParticle2015} For instance, imagine a weather model in which the hypothetical air pressure in a particle varied reasonably within each zone, but a discontinuity at the zone boundary led to a prediction of a tornado forming in the next time step. Note that the algorithm proposed in this paper avoids such discontinuities, but for unrelated reasons can be ill-suited to modeling models with nonlinear dynamics; I will address this issue in later work with a proposed extension to my algorithm.

For the block particle filter method to be useful, covariances between local values must tend to decay with distance. If such decay of correlations holds, then, far enough from a zone boundary, the dynamics return to normal. 

This intuition helps explain the error bounds that these researchers prove their algorithm obeys. They show that the error for the value at a given locus $l$ satisfies

\begin{equation} \label{locerror}
\norm\pi-\hat{\pi}^M\norm_\sell\leq\alpha(\frac{e^{\beta|Z_l|}}{\sqrt{M}}+e^{-\gamma\inf|l-b\in Z_l^{C}|})
\end{equation}

where the constants $\alpha,\beta,\gamma>0$ do not depend on $t$. 
(They define a norm for this purpose which measures the distance between random distributions; I will not reproduce this definition here, as in this this result is merely a guide for intuition.)

One can see that there is a tradeoff: using smaller zones and/or more particles allows better guesses for a given zone to control the term $\frac{e^{\beta| Z_l|}}{\sqrt{M}}$, while using larger zones makes it possible to move away from boundary effects and control the term $e^{-\gamma\inf|m-b\in Z_l^{C}|}$. In practice, Rebeschini and van Handel give a simple example where it would still be possible to control average error per locus, based on finding an optimal balance between adding particles and increasing neighborhood size, but in that example the inverse of their error bound grows only logarithmically with the computing resources/number of particles — in other words, as tolerances tighten, the required number of particles can still grow exponentially. Thus, although their algorithm in practice gives error far lower than that of the bootstrap particle filter, it has not fully overcome the problem of needing exponential computing cost as dimension grows, especially if one wishes to achieve fixed error bounds that are lower than what comes easily from a moderate neighborhood size.

\section{The Finkelstein solution}

In this section, I will sketch out the basic outlines of a recursive algorithm in which each particle at time $t$ is composed of values at different loci which are drawn from state vectors progressed from different particles at time $t-1$. The choice of which values for a given locus combine with which values at other loci is made by running a separate Metropolis-Hastings MCMC to create each composite particle, proposing to replace one locus value at each MCMC step. What acceptance ratio $\rho$ to use for those proposals will be discussed in later sections.

I name this the Finkelstein algorithm, after the character Sally Finkelstein from the movie ``The Nightmare Before Christmas''. I have already compared the block particle filter to a Frankenstein solution, in which the progressed particles are chopped up and then randomly sewn back together. In that algorithm, the suitability of the values in each zone of each particle is measured against observation, but not against the other zones on which it borders. Finkelstein can improve on this. Though herself originally a Frankenstein's-monster-like creation of a stereotypical mad doctor, now that she has been animated, Sally is able to lose her body parts and sew them back on, and thus presumably to choose for herself only those body parts that best fit together. In my terms, Finkelstein would be able to run an MCMC process on her own body, targeting whatever distribution she pleases.

A key aspect of this algorithm is the Metropolis-Hastings acceptance ratio $\rho$. When choosing a formula for such a ratio, the key question is, what distribution do we wish to target? I'll begin by showing an algorithm that targets the natural unnormalized density:

\begin{align} \label{targetdensity}
\begin{split}
f_{\Xtt\sim\tau}(\dens \Xt|\Yt) &=
\int f(\dens \Xt|\Yt, \Xtt)\tau(\Xtt)d\Xtt \\ 
&\propto  \int f(\dens \Yt,\Xt) f(\dens \Xt,\Xtt)\tau(\Xtt)d\Xtt \\ 
&=[\Pi_{l}f(\yt_l|\xt_l)]\int [\Pi_{l}f(\dens z_l|\Xtt)]\tau(\Xtt)d\Xtt
\end{split}
\end{align}

It will turn out that targeting this density still suffers a similar curse of dimensionality as the bootstrap particle filter, so I will modify the algorithm, such that its stationary distribution is not precisely the above expression. Still, this expression is still the starting point; by approximately targeting it, I approximately target $\pi$.

Here are the steps of the basic Finkelstein algorithm. I do not include a formula for the acceptance probability $\rho$ here; I will develop and discuss several alternatives for such a formula, based on modifications of the density above, in the following sections.

\begin{enumerate}
\item Assume we have $\hat\tau^M$, a sampleable distribution which in some sense approximates the ideal filtering distribution $\tau$, and which must be of the form $\frac{1}{M}\sum_{i=1}^M \delta(\Xtt^i)$. Note that unlike the cases of the bootstrap and block particle filters, the final $\hat\pi$ produced by this algorithm already has equally-weighted particles; resampling is not required. 
\item For each particle $\Xtt^i$, progress it to get a full particle $\Raw^i\sim \prog \Xtt^i$ whose local values are known as ${\raw}^i_l$. (In later steps, I will assume for simplicity that there are no duplicate values at any locus, so $i\neq j\Rightarrow{\raw}^i_l\neq {\raw}^j_l$, but relaxing this assumption should be straightforward if necessary.)
\item Find likelihood weights for each such local value, denoted $w^i_l\equiv f(\dens y_l|\raw^i_l)$; and forward densities conditional on $\Xtt^j$ for all $j$ (including $j=i$), denoted $f^{j\rightarrow i}_{l}\equiv f_{\prog }(\dens \raw^i_{l}|\Xtt^j)$.
  
\item In parallel, for $k=1,\dots,M$, do the following:
  
\begin{enumerate}
\item Create a new proposal particle by independently sampling each locus of a vector $\Mcmi^0\in\{1..M\}^L$. This vector specifies the source for the value of $\raw$ that is being considered at each locus; that is, after running the MCMC for $S$ steps, the final value $\Mcmi^S$ will be used to define $\Xt^k$ by setting $\xt_l^k=\raw_l^{\mcmi^S_l}$.  The initial sampling at each locus uses the probabilities
  $$
  P(\mcmi^{0}_l=i)=w^i_l/\sum_{j}w^j_l.
  $$
(Note that these initial sampling probabilities are arbitrary and, if the MCMC successfully runs until convergence, irrelevant. The probabilities above represent a reasonable starting point that should converge reasonably well, but it may be possible to get even faster convergence through some sampling scheme that is not independent across loci.) 
  
\item Run a Metropolis-Hastings MCMC chain targeting an approximation of the filtering distribution, for $s=1,...,S$ steps to (assumed) convergence:

\begin{enumerate}

\item Choose a spatial locus $\lambda(s)\in \{1,\dots,L\}$ uniformly at random. For brevity, I will refer to this as $\lambda$, suppressing the dependency on $s$, in the steps that follow.
\item Sample a proposed particle $\mcmi^*$ from which to draw the replacement value $\raw^{\mcmi^*}_\lambda$ for locus $\lambda$, with probability
  $$
  P(\mcmi^* = i) = w^i_{\lambda}/\sum_j w^j_{\lambda}.
  $$
As with $\lambda$ itself, I am omitting here the subscript $s$, even though this will be resampled at each step. Note that unlike $\Mcmi^s$, which is a vector of one integer per locus, this $\mcmi^*$ is only one integer, which determines the source for the proposed value only at locus $\lambda$. So for convenience in the formulas below, I will also define the vector $\Mcmi^{**}$ such that $\mcmi^{**}_\lambda = \mcmi^{*}$ and $\forall k\in\{1,...,\lambda-1,\lambda+1,...,L\}:\mcmi^{**}_k=\mcmi^{s-1}_k$

\item Accept this proposed change with M-H probability: $1\wedge\rho(\Mcmi^{s-1},\lambda,\mcmi^{*})$, where $\rho$ is defined below. In case of acceptance, $\Mcmi^{s}:=\Mcmi^{**}$; Otherwise, in case of rejection, make no change, so that $\Mcmi^{s}:=\Mcmi^{s-1}$.

\end{enumerate}
\item Let $\xt_l^k=\raw_l^{\mcmi^S_l}$.

\end{enumerate}
\end{enumerate}

We would like to tune the proposal and acceptance probabilities so that this algorithm targets the distribution $(\Xt|\Yt,\{\Xtt^i\})$. Insofar as we succeed, the full algorithm will be very similar to a bootstrap particle filter, which similarly targets that distribution. Thus, on a purely intuitive level, it is unsurprising that this should work if the MCMC does. But the whole process depends on the validity of the MCMC, which in turn depends on the M-H acceptance probability $\rho(\Mcmi^{s-1},\lambda,\mcmi^{*})$, left unspecified above.

\section{Developing a working formula for \texorpdfstring{$\rho$}{rho}}

In this section, I'll first develop some motivating understanding of what $\rho$ should be like, then develop two specific formulas for $\rho$:

\begin{enumerate}

\item The first formula, $\rho_\mathrm{full}$, is for illustrative purposes only. Though it is, by construction, asymptotically correct--that is, the algorithm using $\rho_\mathrm{full}$ approaches the correct filtering distribution as the number of particles $M$ approaches infinity--it is unsuitable for use in practice. Not only does it lead to impractically high computational costs for a specific $M$, it also suffers from similar dimensionality problems as the bootstrap particle filter.
  
\item The second formula $\rho_\mathrm{local}$ only considers values of the MCMC in some local neighborhood of $\lambda(s)$. It thus does not suffer the weight degeneracy problem of the bootstrap particle filter or $\rho_\mathrm{full}$, while still being possible to calculate in computing time that's polynomial in number of particles, and linear in dimension and number of time steps.

\end{enumerate}

In the next section, I'll develop a further formula for $\rho$ that improves the computational characteristics, as well as proposing some other computational optimizations.
            
\subsection{\texorpdfstring{$\rho_\mathrm{full}$}{rho\_full}: full acceptance probability}

I will begin by using a standard Metropolis-Hastings ratio to derive a $\rho_\mathrm{full}$ that targets the (unnormalized) density~\ref{targetdensity}. That expression is promising in one sense: it suggests that one can judge the fit of local proposals drawn from two different particles using an expression involving the transition kernel forward density. However, the fact that it involves a sum over all particles, of a product over all loci, makes $\rho_\mathrm{full}$ computationally unworkable in practice. And even if there were enough available computational power to calculate this at every step of an MCMC, taking a product over loci would lead to similar problems as the naive high-dimensional bootstrap particle filter has.

If we propose replacement $\xt^i_l$ for the value at a given locus $l$ with probability proportional to some value $\weight^i_l$,
we can construct a Metropolis-Hastings acceptance probability in the usual way, by multiplying a ratio of target densities (proposed over current) by a ratio of proposal densities (current over proposed). 

(To avoid nested subscripts/superscripts in the following, I leave out redundant indices for locus; thus using the notation $f^{j\rightarrow \mcmi^{**}}_{k}$ rather than $f^{j\rightarrow \mcmi^{**}_k}_{k}$, using $f^{j\rightarrow \mcmi(s-1)}_{k}$ rather than $f^{j\rightarrow \mcmi^{s-1}_k}_{k}$, and using $w^{\mcmi(s-1)}_k$ rather than $w^{\mcmi^{s-1}_k}_k$.)

\begin{equation} \label{puremhacceptanceratio}
\rho_\mathrm{full}\equiv\frac{
[{w^{\mcmi^*}_\lambda}{\Pi_{k\neq \lambda} w^{\mcmi(s-1)}_k}]
~\Sigma_{j=1}^M[\Pi_{k}f^{j\rightarrow \mcmi^{**}}_{k}]
}
{
[{w^{\mcmi(s-1)}_\lambda}{\Pi_{k\neq \lambda}w^{\mcmi(s-1)}_k}]
~\Sigma_{j=1}^M[\Pi_{k}f^{j\rightarrow \mcmi(s-1)}_{k}]
}
\cdot \frac{
\weight^{\mcmi(s-1)}_\lambda~\Sigma_{j=1}^M f^{j\rightarrow \mcmi(s-1)}_\lambda
}
{
\weight^{\mcmi^{**}}_\lambda~\Sigma_{j=1}^M f^{j\rightarrow \mcmi{**}}_\lambda
}
\end{equation}

Let's consider these terms from left to right, taking the version of each term as it appears in the numerator:

\begin{enumerate}

  \item ${w^{\mcmi^*}_\lambda}$: The likelihood at the locus in question; the term which depends on $\Yt$. I will set $\weight^{\mcmi^*}_\lambda$ so as to cancel this out.
  
  \item ${\Pi_{k\neq \lambda}w^{\mcmi(s-1)}_k}$: The likelihoods at other loci. These cancel out naturally.
  
  \item $\Sigma_{j}[\Pi_{k}f^{j\rightarrow \mcmi^{**}}_{k}]$: This sum of products term is the heart of the calculation at each MCMC step. Each product is the forward likelihood of the current/proposed hybrid particle conditional on a given history; the sum of products is proportional to the forward likelihood of the current/proposed hybrid particle conditional on $\hat\tau^M$.
  
  \item $\weight^{\mcmi(s-1)}_\lambda$: Weights that define the proposal distribution and can be chosen arbitrarily (up to normalization).
  
  \item $\Sigma_{j=1}^M f^{j\rightarrow \mcmi(s-1)}_\lambda$: The part of the proposal density that's due to $\hat\tau^M$; the probability density that a given value would have been in $\{\Raw^i_\lambda:0<l<L\}$ to be available to be sampled. To someone used to other variations of particle filtering algorithms, it may seem counterintuitive to include this term; usually, taking advantage of the forward density is a key aspect of how the algorithm works, not something that needs canceling out. However, from the perspective of a Metropolis-Hastings construction, it is necessary to include this in order to target the intended (unnormalized) density~\ref{targetdensity}. On a more intuitive level, one might note that the forward density values $f^{j\rightarrow \mcmi(s-1)}_\lambda$ for each $j$ appear both here and in the sum of products term in the denominator; so if one did not include this term, that would in a sense be double counting these forward density values by allowing them to cause an increased proposal density and then also increase the acceptance ratio.

  \end{enumerate}

The unnormalized proposal weights $\weight^i_l$ can be set at will; I will let $\weight^i_l\equiv w^i_l$, so that these terms cancel out. Now $\rho$ is just a ratio of sums of products, multiplied by a ratio of the forward mean proposal densities  $\bar{f}^{i}_{l}$. I could have set $\weight^i_l$ to also cancel out $\bar{f}^{i}_{l}$, but as seen later, this would slow convergence. 

The quantities $\bar{f}^{i}_{l}$ do not change for different MCMC chains or for different steps in each chain and can be precalculated in $O(LM)$ time. We thus have 

\begin{align} \label{simplifiedmhacceptanceratio}
\rho_\mathrm{full}&\equiv\frac{\Sigma_{j=1}^M[\Pi_{k=1}^{L}f^{j\rightarrow \mcmi^{**}}_k]
}{\Sigma_{j=1}^M[\Pi_{k=1}^{L}f^{j\rightarrow \mcmi(s-1)}_{k}]} 
\frac{\Sigma_{j=1}^M f^{j\rightarrow \mcmi(s-1)}_l
}
{
\Sigma_{j=1}^M f^{j\rightarrow \mcmi^{**}}_l
}
\end{align}

By construction, this acceptance probability satisfies the conditions for detailed balance of a Metropolis-Hastings MCMC, and thus converges to the desired target stationary distribution \ref{targetdensity} under standard regularity assumptions. At convergence, the algorithm will give $M$ samples from the correct target distribution, but with the constraint that the value for each sample at each locus must be available in $\tilde{\Xt}^{1..M}$. As $M\rightarrow\infty$, the set of values available at each locus will become dense, so the conditionality will not be restrictive. Thus, asymptotically, one would expect that each $\Xt^i$ should be a sample from the correct filtering distribution, conditional on $\Xtt\sim\hat\tau^{M_{t-1}}$.

Yet $\rho_\mathrm{full}$ is not useful in practice, for two reasons. First, on a relatively trivial level, actually calculating the sum of products terms once for each step of the MCMC would be computationally prohibitive; though not exponential, the resources required would be extreme. 

More importantly, unless $M_{t-1}$ is exponentially high, $\cond _t\prog \hat\tau^{M_{t-1}}$ is not a good approximation of $\cond _t\prog \tau$, because of a curse of dimensionality very similar to that which causes the bootstrap particle filter to fail in high dimensions. In $\rho_\mathrm{full}$, the acceptance probability is based on a sum over histories of products over loci of likelihoods. Following a similar logic as Bickel \etal\cite{bickelSharpFailureRates2008}, discussed above, for showing weight degeneracy in the bootstrap particle filter, we see that, assuming that the likelihoods associated with distantly-separated loci are roughly independent, then as dimension increases the distribution of these products over loci will approach a log-normal. Since the variance of that distribution will grow with dimension, the sum is likely to be degenerate unless number of particles grows exponentially with dimension; just one history particle will contribute more to the sum than all others put together.

Consider the example of a weather model of the continental United States, where imperfect measurements of atmospheric conditions are taken daily over a set of cities. In this case, the sum would be degenerate because, although any given proposal particle (weather map for today) might accord well with a given history (possible weather map for yesterday) for some cities, you'd nevertheless need to consider an exponentially large number of possible histories before finding one which accords well across \textit{all} cities with a given realistic present.

In essence, rather than resolving the high-dimensionality problem at time $t$, we've merely pushed it off to time $t-1$; because of the high variance of the product terms $\Pi_{k=1}^{L}f^{j\rightarrow \mcmi^{**}}_{k}$, the sums will tend to be dominated by the product term for a single history $j$, losing most of the benefits of a high number of particles.

\subsection{\texorpdfstring{$\rho_\mathrm{local}$}{rho\_local}: Simplifying the product terms by focusing on local neighborhoods}

The main computational burden of calculating $\rho_\mathrm{full}$ comes from the sum over histories of a product over loci. To deal with these computational issues, as well as with the degeneracy of the sum, it would be good to take this product over fewer terms. To do so, I restrict the product over loci to only consider loci in some neighborhood of the locus $l$ which the proposal would change. 

This idea gains some support from the decay of correlations property of that Rebeschini and van Handel (2015) demonstrate. This is a complex issue which occupies a significant portion of their paper, but to summarize briefly: they assume particle filters with local dynamics, and both forward densities and observation likelihoods that are strongly bounded away from zero and infinity, Given those assumptions (which they argue are probably stronger than necessary in most practical cases), they show that changing the value at locus $k$ cannot change the conditional distribution of the value at $l$ by more than a quantity that falls exponentially as the distance between $k$ and $l$ increases. Thus, it would seem logical that, in calculating an acceptance probability to target the distribution at $l$, one may safely ignore faraway loci $k$.

To use this idea for the Finkelstein algorithm, assume there is a natural distance metric $d(l,k)$ over loci; for instance, if loci were arranged in a square lattice, $d(l,k)$ could be the $\ell_1$-distance. Use this distance to define neighborhood balls $\ball_r(\lambda)\equiv\{l:d(\lambda,l)\leq r\}$, and use the natural notation that $\Xtt_{\ball_r(\lambda)}\equiv\{x_l:l\in\ball_r(\lambda)\}$. Thus, the new $\rho$ would be:

\begin{equation} \label{localacceptanceratio}
\rho_\mathrm{local}\equiv\frac{\Sigma_{j=1}^M[\Pi_{k\in \ball_r(\lambda)}f^{j\rightarrow \mcmi^{**}}_k]
}{\Sigma_{j=1^M}[\Pi_{k\in \ball_r(\lambda)}f^{j\rightarrow \mcmi(s-1)}_{k}]}
\frac{\Sigma_{j=1}^M f^{j\rightarrow \mcmi(s-1)}_\lambda
}
{
\Sigma_{j=1}^M f^{j\rightarrow \mcmi^{*}}_\lambda
}
\end{equation}

If this works, it will have finally conquered the curse of dimensions. For any locus $l$, there are only $|\ball_r(l)|$ terms in each product; a quantity which does not depend on the overall dimension of the problem, only on the local connectivity. We can therefore choose a fixed $M$ large enough such that this sum of products is not degenerate (not dominated by just one of the products) for $N\equiv \max_{l}|\ball_r(l)|$ loci.

However, it should be noted that with this acceptance probability, the algorithm is no longer strictly speaking Metropolis-Hastings. In particular, the overall MCMC is no longer guaranteed to obey detailed balance. If one repeatedly replaced the values of a single locus $l$, the MCMC would, by the standard Metropolis-Hastings construction, show detailed balance at a unique stationary distribution with a density proportional to:

\begin{equation} \label{localdensity}
f_l(\xt_l|\xt_1,...,\xt_{l-1},\xt_{l+1},...,\xt_L)\equiv(\Pi_{\lambda=1}^L\mathds{1}_{\xt_\lambda\in\{\raw_\lambda\}})\Sigma_{j=1}^M[\Pi_{k\in \ball_r(l)}f_{\prog }(\dens \xt_{k}|\Xtt^j)]
\end{equation}

This function is  not only of $\xt_l$, but of all $\xt_k$ such that $k\in\ball_r(l)$. However, since this density is not the same for two different values of $l$, this detailed balance can and will break down. The MCMC is still uniformly ergodic, so a unique stationary distribution still exists; but without detailed balance, we lack the nice guarantees that Metropolis-Hastings would offer as to what that target distribution is. At present, then, my use of $\rho_\mathrm{local}$, and all later versions of $\rho$ that build on it, is based on empirical validation, as seen below in the simulation section, not rigorous theory.

\section{Computational optimizations}

\subsection{\texorpdfstring{$\rho_\mathrm{sampled}$}{rho\_sampled}: a version of \texorpdfstring{$\rho_\mathrm{local}$}{rho\_local} which replaces numerator and denominator by unbiased estimators}

Running the Finkelstein algorithm with acceptance probability $\rho_\mathrm{local}$ does not require exponential computation, but even polynomial amounts of computation can be daunting in practice. Recall that the of sums over all histories in $\rho_\mathrm{local}$ are proportional to the forward likelihood conditional on $\hat\tau^M$, the $M$-particle approximation of the filtering distribution $\tau$. At each step, we can save computation by, effectively, estimating $\tau$ with only an arbitrary fixed number $H<<M$ particles; that is, by using only $H$ history terms for these sums and using those to get unbiased Horvitz-Thompson estimators of the totals. This leads to $\rho_\mathrm{sampled}$. Note that the specific $H$ history particles used will change from step to step and locus to locus in the MCMC, thus taking advantage of the full $M$ particles in equilibrium.

The idea of using unbiased estimators to calculate the Metropolis-Hastings acceptance ratio is not new, and, as Andrieu and Roberts 2009\cite{andrieuPseudomarginalApproachEfficient2009a} show, this can be made to conserve the stationary distribution, provided that any randomness used in finding the estimator is maintained as part of an expanded Metropolis-Hastings parameter space. This could work for $\rho_\mathrm{full}$. But now that we are working from $\rho_\mathrm{local}$, this is impossible because the MCMC is already not true Metropolis-Hastings with a single common parameter space. As discussed above, the target distribution of the particle as a whole is different when changing values at different loci, although there's reason to hope that the difference for nearby loci is small. Thus, $\rho_\mathrm{sampled}$ will inevitably have a different stationary distribution from $\rho_\mathrm{local}$. Nevertheless, in the algorithm below, in an attempt to ensure that the stationary distribution changes as little as possible, I expand the parameter space of $\rho_\mathrm{sampled}$ with a matrix $\Hist^s$. This ensures that the same $H$ histories used when accepting a value at a locus are also used when deciding whether to change that value later.

We revise the algorithm from section 2 as follows:

\begin{enumerate}
\item As before, assume we have $\hat\tau^M$.
\item As before, for each particle $\Xtt^i$, progress it to get a full particle $\Raw^i\sim \prog \Xtt^i$.
\item As before, find likelihood weights $w^i_l\equiv f(\dens y_l|\xt^i_l)$ and forward densities $f^{j\rightarrow i}_{l}\equiv f_{\prog }(\dens \xt^i_{l}|\Xtt^j)$ for all $i,j\in\{1,\dots,M\}$ and $l\in\{1,\dots,L\}$. 
  
\item In parallel, for $k=1,\dots,M$, do the following:
  
\begin{enumerate}
\item As before, sample $\Mcmi^0\in\{1..M\}^L$ to initialize the state of the new proposal particle.

\item In addition, initialize an $L\times H$ matrix $\Hist^0$ with entries in $\{1...M\}$, sampled iid with probabilities 
$$
P(\hist_{l,h}^0=i) =
\frac{g(f^{i\rightarrow \mcmi^0_l}_l)}
{\Sigma_j g(f^{j\rightarrow \mcmi^0_l}_l)},
$$
where $g$ is an arbitrary monotonically increasing function.

Each entry $\hist_{l,h}^0$ gives the index $i$ of one history $\Xtt^i$ which we will later use to estimate the denominator of $\rho$. The significance of $g$ will be explained in more detail below.

\item As before, run a Metropolis-Hastings MCMC chain, for steps $s=1,...,S$, updating $\Hist$ and $\Mcmi$ at each step:

\begin{enumerate}

\item As before, choose a spatial locus $\lambda(s)$ (aka $\lambda$) uniformly at random.
\item As before, sample a proposed replacement $\mcmi^*(s)$ (aka $\mcmi^*$) for locus $\lambda$, with probability 
$$
P(\mcmi^* = i) = w^i_{\lambda}/\sum_j w^j_{\lambda}.
$$
Also, define $\Mcmi^{**}$ as before.
\item In addition, sample (iid) a set $(\hist^*_1,...,\hist^*_H)$ of histories by which to judge this proposed replacement, where 
$$
P(\hist^*_{h}=i) =
\frac{g(f^{i\rightarrow \mcmi^*}_\lambda)}
{\Sigma_j g(f^{j\rightarrow \mcmi^*}_\lambda)}.
$$
Define the matrix $\Hist^{**}$ by 

\[
    \hist^{**}_{l,h}:= 
\begin{dcases}
    \hist^{*}_{h}              & \text{if } l=\lambda\\
    \hist^{s-1}_{l,h}& \text{if } l\neq\lambda
\end{dcases}
\]

\item Finally, define
\begin{equation} \label{rhosampled}
\rho_\mathrm{sampled}\equiv
\frac{\Sigma_{h\in\{1..H\}}\frac{1}{g_\lambda(\hist^*_h,\mcmi^{**})}[\Pi_{l\in\ball_r(\lambda)}f^{\hist^*_h\rightarrow \mcmi^{**}}_{l}]}
{\Sigma_{h\in\{1..H\}}\frac{1}{g_\lambda(\hist^*_h,\mcmi^{s-1})}[\Pi_{l\in\ball_r(\lambda)}f^{\hist^*_h\rightarrow \mcmi^{s-1}}_{l}]}
\frac{\Sigma_{j=1}^M f^{j\rightarrow \mcmi^{s-1}}_\lambda
}
{
\Sigma_{j=1}^M f^{j\rightarrow \mcmi^{**}}_\lambda
},
\end{equation} 
where 
$$
g_l(i,j)\equiv\frac{g(f^{i\rightarrow j}_l)}
{\Sigma_k g(f^{k\rightarrow j}_l)}.
$$
As usual, we accept the proposed replacement with M-H probability $1\wedge\rho_\mathrm{sampled}$. In case of acceptance, let $\Mcmi^{s}:=\Mcmi^{**}$ and $\Hist^s:=\Hist^{**}$. Otherwise, make no change, so that $\Mcmi^{s}:=\Mcmi^{s-1}$ and $\Hist^s:=\Hist^{s-1}$.

\end{enumerate}
\item As before, set $\xt_l^k=\raw_l^{\mcmi^S_l}$.

\end{enumerate}
\end{enumerate}

In addition to showing that using unbiased estimators can conserve the target distribution when the parameter space is expanded, Andrieu and Roberts also discuss the case where the parameter space is not expanded. They show that this case still has a stationary distribution, which converges to the original target distribution as more samples are taken. With minor modifications, the same proof applies to the MCMC using $\rho_\mathrm{sampled}$; the stationary distribution of $\rho_\mathrm{sampled}$ is not the same as that of $\rho_\mathrm{local}$, but converges to it as $H\rightarrow\infty$.

A few words on the function $g$. First, note that the sampling probabilities for $\Hist^s$ are in principle arbitrary, so could be a function of the full current vector $\Mcmi^s$. In the above discussion, however, they are a function of only $f^{i\rightarrow\mcmi^*}_l$; this simplifies both notation and computation. Second, $g$ should be chosen to be some non-decreasing function of $f^{i\rightarrow\mcmi^*}_\lambda$, so that the variance in $\Pi_{l\in\ball_r(\lambda)}f^{\hist^*_h\rightarrow \mcmi^{**}}_{l}$ is at least partially offset by that in $g$, increasing the efficiency of the estimation process. Another way of saying this is that we should make it more likely to sample plausible histories than implausible ones. Possible choices for $g$ are discussed in Section \ref{historyweights} below. Whatever $g$ is chosen, the denominator ${\Sigma_j g(f^{j\rightarrow \mcmi^*}_\lambda)}$ can be precalculated for each possible choice of $\mcmi^*$, meaning that this does not meaningfully increase computing requirements per MCMC step.

How much computation does $\rho_\mathrm{sampled}$ save? The $\rho_\mathrm{local}$ algorithm requires running $M$ different MCMC chains, with each of $L$ loci going through $S$ steps, and at each step calculating a $\rho$ using a sum over $M$ histories of a product over the up to $N$ locations in the relevant $\ball_r(l)$. The total computation cost is at least $O(M^2LNS)$. This is better than $O(M^2L^2S)$ that $\rho_\mathrm{full}$ would have taken, but still somewhat burdensome. To get $\rho_\mathrm{sampled}$, on the other hand, we only calculate the product of locus likelihoods for an arbitrary number $H$ of histories rather than all $M$ of them. Thus, the total computation cost falls to $O(MHLNS)$; since $H$ and $N$ are arbitrary constants that can be set independently from the full size $M$ and $L$ respectively, this is a substantial improvement.

\subsection{Discussion of proposal weights}

In the above discussion, the proposal weights $\weight_l^i$, used by all versions of $\rho$, are arbitrary. That is, any proposals with nonzero weights could be used; the $\weight_l^i$ are accounted for out in the $\bar{F}$ term.

Ideally, these proposal weights should both be tuned for maximum efficiency of the MCMC; that is, insofar as it does not substantially increase the computational costs per step, to try to ensure that the variance of the acceptance probability is as low as possible (approaching the Gibbs sampling case where it's uniformly 1) while maintaining disperse (high-variance) proposal values for good ergodicity/mixing.                         

For $\weight_l^i$, that is similar to the idea of an optimal proposal distribution, which is common in the particle filter literature.\cite{snyderParticleFiltersOptimal2011} The low-dimensional bootstrap particle filter uses $(\xt|\xtt^{1..M})$ as a proposal, then reweights using $(\yt|\xt)$. In such a case, the idea of an ideal proposal distribution is that if you could propose from $(\xt|\xtt^{1..M},y)$, the reweighting step would not be necessary. 

Applying a similar idea to $\weight_l^i$, it becomes clear why I have set it equal to $w_l^i$. Of course, including a factor of $w_l^i$ in $\weight_l^i$ helps these terms cancel and thus simplifies the calculation of $\rho$. But if simplicity of calculating $\rho$ were the only consideration, I could have set $\weight_l^i$ to $w_l^i{\Sigma_{j=1}^M f^{j\rightarrow i}_l}$, so that the ratio $\frac{\Sigma_{j=1}^M f^{j\rightarrow \mcmi(s-1)}_l}{\Sigma_{j=1}^M f^{j\rightarrow \mcmi(s)*}_l}$ would cancel out too.

But setting $\weight_l^i\equiv w_l^i$ ensures that the proposal density for $\xt_l$, conditional on $\Xtt^{1..M}$ and $\Yt$, is $(\xt_l|\Xtt^{1..M},\yt_l)$ — not too far from the ideal $(\xt_l|\Xtt^{1..M},\Yt,\Mcmc^{s-1}_{\ball_r(l)\setminus l})$ which would allow an acceptance probability of uniformly 1. That's because the density of $(\xt_l|\Xtt^{1..M})$ is included implicitly through the progression procedure, while that of $(\xt_l|\yt_l)$ is handled explicitly through $w_l^i$.

\subsection{Refining the history sampling weights} \label{historyweights}

What about the history sampling weights $g_l^{i\rightarrow j}$? As above, these are arbitrary. Ideally, to minimize the variance of the acceptance probability, they would approximate: 

\begin{align} \label{idealg}
g_l^{i\rightarrow j}(\Mcmc^{s-1}_{\ball_r(l)\setminus l})&\propto f_{\prog }(\raw^j_l|\Xtt^i,\Mcmc^{s-1}_{\ball_r(l)})\\
&=\Pi_{\kappa\in\ball_r(l)}f^{i\rightarrow j}_{\kappa }
\end{align}

... because this expression in the Horvitz-Thompson inverse sampling weight term would cancel exactly with its relative contribution to the estimated sum of procucts, so that the overall estimator would be governed solely by the sum of weights term in the denominator.

It is computationally infeasible to calculate these quantities exactly for each step of the MCMC, so I use $g_l^{i\rightarrow j}$. In the simulation below, I've tested two formulas for these weights: 

\begin{enumerate}
  \item $g_\mathrm{uniform}(x)=1$
  
  \item $g_\mathrm{bentlog}(x)=\frac{\log(f^{i\rightarrow j}_\lambda)-\log(\min(f^{i\rightarrow j}_\lambda)}{\bla} + max(0,\log(x)-\log(\max(x)+\blb)$, where $\min(x)$ and $\max(x)$ are the precalculated minimum and maximum values of $f^{i\rightarrow j}_\lambda$ and $\bla, \blb$ are positive constants. 
\end{enumerate}

Both of these options are simply computationally-convenient first attempts; though simulations show $g_\mathrm{bentlog}$ is an improvement over $g_\mathrm{uniform}$, it is surely not optimal in this regard. In further work, I will look into using proposal distributions that are conditional on the current values at other loci, not just on the observations at the current locus.

\subsection{Theoretical limitations of the algorithms in this paper}

Both the block particle filter and the Finkelstein particle filter proposed here are intended to deal with the curse of dimensionality. However, both may fail in cases where forward densities — that is, the relative probabilities of given states at time $t$ conditional on the state at time $t-1$ — are concentrated around particular values, and thus insufficiently ergodic. Rebeschini and Van Handel's error bounds rely on a strong ergodicity assumption, bounding the forward density away from $0$ in a way that they themselves acknowledge is unrealistic in real-world cases. In a separate paper, they explore further the kind of problems that can arise when this assumption does not apply, and the regime where that failure occurs in practice.\cite{rebeschiniPhaseTransitionsNonlinear2015}

For the Finkelstein particle filter, I am not giving any formal proofs of performance, but it is clear that if ergodicity is poor enough, my algorithm will also break down. For example, suppose that the forward density from history $\Xtt^j$ to raw locus value $\xt^i_l$ is less than $\epsilon$ if $i\neq j$, and otherwise greater than $100\epsilon$. In that case, the MCMC will strongly tend to get stuck in states where all locus values come from the same history. Metaphorically, Sally Finkelstein would be too picky, never selecting nearby body parts that didn't match perfectly. The result would then reduce to the bootstrap particle filter, with more useless computational cost.

Do the real-world problems to which these algorithms are applied have enough ergodicity for them to function? For these algorithms to be appropriate, we'd need a situation with enough nonlinear effects that simple Kalman filters don't suffice; yet also one which still has plenty of new randomness at each time step, such that even if the forward density is not actually strongly ergodic, it is at least diffuse enough for these algorithms to work. In SLAM (simultaneous location and mapping) models for robotics applications, such situations arise. But in fluid dynamics models, chaotic dynamics are the rule. Such models can be deterministic or nearly so, with highly concentrated forward densities, yet still have interesting dynamics. Uncertainty in the initial conditions is amplified at each time step, so even in a deterministic model with new measurement tending to reduce the uncertainty at each time step, overall uncertainty can remain in equilibrium.   

Due to the ``picky Sally'' problem explained just above, the Finkelstein algorithm as explained in current paper does not deal well with such deterministic or nearly-deterministic models. However, in a follow-up paper, I will offer a modification of this algorithm to address such models.

\section{Numerical simulations}

\subsection{Setup}

Filtering algorithms cannot be expected to precisely infer the underlying true state of the hidden Markov model. Instead, the goal is merely to infer its conditional distribution, and most of the interest of the problem lies in the fact that this distribution remains non-degenerate; as we acquire more information in order to narrow down the possible states, the state itself evolves, so we never catch up. 

Thus, we cannot simply follow the recipe of a more traditional simulation study of a technique for parameter inference. In traditional parameter inference, the object of interest is the true parameter value(s), which can be arbitrarily chosen when running a simulation. Although inference algorithms may yield an inferred distribution for the parameter(s), the interpretation of this distribution as a confidence distribution (for frequentist methods) or a credible distribution (for Bayesian methods) is in some sense not inherent to the problem; for example, in the case of Bayesian methods, a credible interval is only as valid as the priors that produce it. However, in this case, we are not making arbitrary assumptions in order to get the best or the most robust performance; the assumptions are given by the problem, and the aim is to calculate a true probability distribution. In order to efficiently measure an algorithm's performance, we'd like a setting where the correct value of the object of interest -- not the point value, but the conditional distribution -- is known.

So, in order to run a simulation study, I fall back on a linear Gaussian model, where the Kalman filter algorithm gives an analytically-correct filtering distribution. Of course, given that such an analytic solution does exist, one would never in practice use an inexact filtering algorithm such as those discussed by this paper. However, the ability of our more-general algorithm to roughly reproduce the results of a Kalman filter is, at the least, encouraging.

The particular linear Gaussian model I use is a model based on a progression matrix $P$, a novelty matrix $N$, and a measurement error covariance matrix $E$:

\begin{align} \label{kfdef1}
z &=Px + \delta \\
y &=z + \epsilon \\
\delta &\sim \mathcal{N}(0,N)  \\
\epsilon &\sim \mathcal{N}(0,E)
\end{align}

$P$ is a tridiagonal matrix. $N$ and $E$ are diagonal matrices with periodic structure, so that some loci are best learned about through their neighbors. Specifically, $N$'s elements alternate between a higher and a lower variance, each value occurring at every 2nd locus, while $E$ has a lower variance at every 5th locus.:

\begin{align} \label{kfdef2}
P &\equiv \begin{bmatrix}
   {\kfb} & {\kfc} & {   } & {   } & { 0 } \\
   {\kfa} & {\kfb} & {\kfc} & {   } & {   } \\
   {   } & {\kfa} & {\kfb} & \ddots & {   } \\
   {   } & {   } & \ddots & \ddots & {\kfc}\\
   { 0 } & {   } & {   } & {\kfa} & {\kfb}\\
\end{bmatrix} \\
\kfa &\equiv .4; \kfb\equiv.35; \kfc\equiv.05; \kfa+\kfb+\kfc=.8<1 \\
N &\equiv \begin{bmatrix}
   { 1 } & {   } & {   } & {   } & {   } & { 0 } \\
   {   } & {\kfd}& {   } & {   } & {   } & {   } \\
   {   } & {   } & { 1 } & {   } & {   } & {   } \\
   {   } & {   } & {   } & {\kfd}&       & {   } \\
   {   } & {   } & {   } &       & \ddots& {   }\\
   { 0 } & {   } & {   } & {   } & {   } & {\kfd}\\
\end{bmatrix} \\
\kfd &= .25 \\
\renewcommand{\kbldelim}{[}
\renewcommand{\kbrdelim}{]}
E &\equiv \kbordermatrix{
         & { 1 } & { 2 } & \dots & { 4 } & { 5 } & { 6 } & \dots & { d } \\
   { 1 } & \kfe  & {   } & {   } & {   } & {   } & {   } & {   } & { 0 } \\
   { 2 } & {   } & { 1 } & {   } & {   } & {   } & {   } & {   } & {   } \\
   \vdots& {   } & {   } & \ddots& {   } & {   } & {   } & {   } & {   } \\
   { 4 } & {   } & {   } & {   } & { 1 } & {   } & {   } & {   } & {   }\\
   { 5 } & {   } & {   } & {   } & {   } & \kfe  & {   } & {   } & {   }\\
   { 6 } & {   } & {   } & {   } & {   } &  {   }& { 1 } & {   } & {   }\\
   \vdots& {   } & {   } & {   } & {   } & {   } & {   } & \ddots& {   } \\
   { d } & { 0 } & {   } & {   } & {   } & {   } & {   } & {   } & { 1 }\\
} \\
\kfe &= .16 \\
\end{align}

The state was initialized at mean 0 and variance 5 independently at each locus. The model was run for 10 time steps, and for the particle filtering algorithms the outcome variables of 5 separate runs were averaged.

A number of parameters were tried for the algorithms, but a good set of numbers for comparing different models was: $400$ particles for the Finkelstein variants, $400^2=160000$ particles for the bootstrap particle filter, and $400^2/5=32000$ particles for the block particle filter algorithm. These numbers were chosen so that each algorithm would take roughly comparable computing time; the only step that requires computing power that is quadratic in the number of particles is precalculating the forward densities $f^{j\rightarrow i}_{l}$ in the Finkelstein algorithm.

All results for Figures \ref{sqerrplot} and \ref{stabilityplot} are for a 30-dimensional model. Results for both Finkelstein and block particle filter algorithms remained materially similar as model dimension was varied from 30 to 90, demonstrating that the Finkelstein and Frankenstein algorithms' errors are roughly independent of dimension, as expected.

The Finkelstein algorithm was used with $\rho_\mathrm{sampled}$, with 45 histories per location and two formulas for the history sampling probabilities $g$ ($g_\mathrm{uniform}$, and $g_\mathrm{bentlog}$ with $\bla=\blb=5$).  The neighborhood width was $r\equiv 1$, which is to say that $\max |\ball_r(l)|=3$. Similarly, the zone size for the block particle filter algorithm was 3. 

Note that I am not the first to simulate outcomes for the block particle filter. Although Rebeschini and van Handel's paper originally proposing it relied on proofs rather than simulations, more recent papers have implemented it and given results.\cite{morzfeldVariationalParticleSmoothers2018} \cite{farchiReviewArticleComparison2018} \cite{yaxianSequentialMonteCarlo2018} The results there are not directly comparable with those given here due to different models used.

\subsection{Results}

\begin{figure} 
\includegraphics[scale=1]{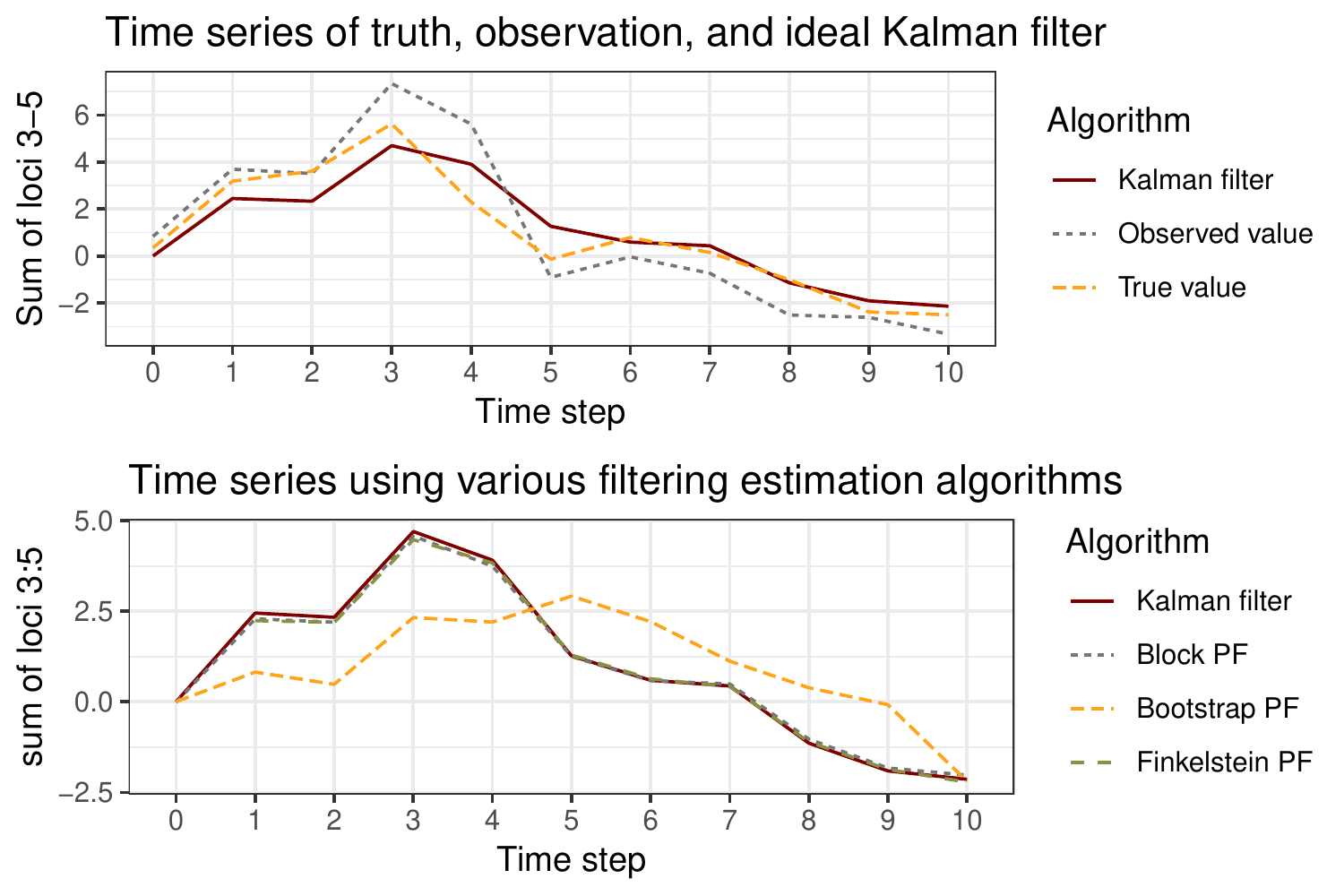}
\caption[]{Time series from a single run of each algorithm for 10 time steps, in a model with 90 dimensions. Parameters for each algorithm are given in text.}
\label{trueevol}
\end{figure}

To get an intuition for this situation, I will begin by showing the evolution of a single run of the model. The top panel of figure~\ref{trueevol} shows the evolution over time of the true value ($\sum_{l=3}^5\xt_l$), the observed value ($\sum_{l=3}^5\yt_l$), and the filtering distribution as calculated by a Kalman filter ($\E_\pi(\sum_{l=3}^5\xt_l|\Yt_1,...,\Yt_t$), for the sum of loci 3-5. The bottom panel shows the the mean of the estimated filtering distribution for each of the four algorithms I tested — Kalman filter (analytically correct), bootstrap particle filter, block particle filter, and Finkelstein particle filter.

In the upper panel of Figure~\ref{trueevol}, one can see that the observations vary relatively widely around the truth, while the Kalman filter mean follows those observations with more conservative moves, thus staying closer to the true value. In the lower panel, one can see that the bootstrap particle filter falls to the curse of dimensionality; while the block and Finkelstein particle filters both manage to approximate the correct Kalman filter distribution relatively well.

\begin{figure}
\includegraphics[scale=1]{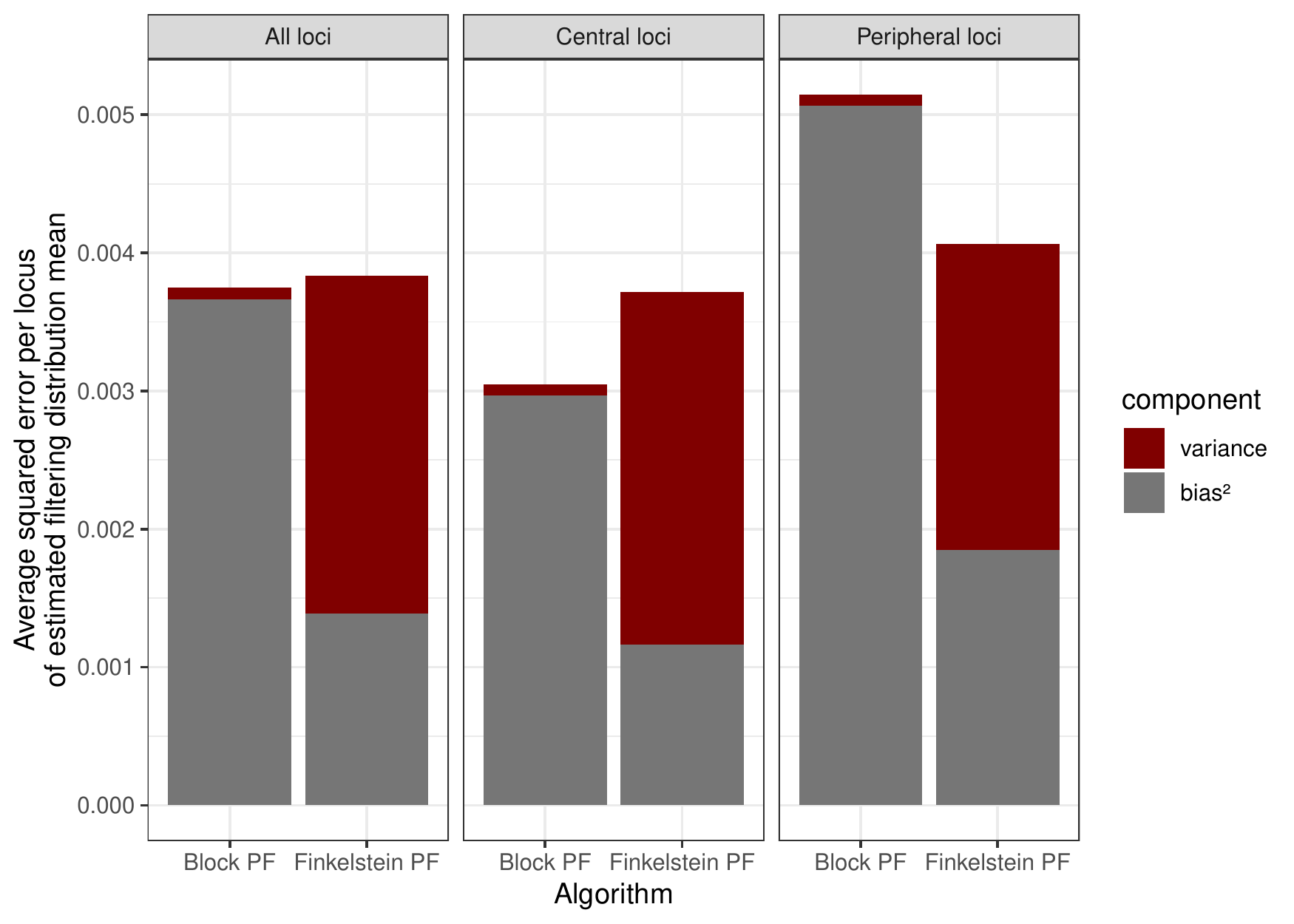}
\caption[]{Breakdown of average squared error per locus.}
\label{sqerrplot}
\end{figure}

To compare outcomes of the two working algorithms in greater depth, Figure~\ref{sqerrplot} shows the average squared error per locus: that is, the squared difference between the estimated distribution mean and the correct mean as given by the Kalman filter, conditional on a single fixed series of observations $(\yt_1,...,\yt_{10})$. Note that we are measuring error relative to the mean of the Kalman filter rather than to $\xt_l$; this is because the Kalman filter result is the ideal filtering distribution that we are trying to capture here. Though it's not visible in these graphs, both algorithms do a relatively good job of reproducing the variance of the filtering distribution; this is within 3\% of the true value for both algorithms.

Figure~\ref{sqerrplot} makes two things clear. First, there is a bias/variance tradeoff between the block particle filter and the Finkelstein particle filter; with comparable run times, the block filter has a higher bias but almost no variance in its distribution error. Second, as expected, the performance of the block particle filter differs for loci that are central to their neighborhood $Z_j$, as opposed to loci that are on the border of their neighborhood. (The small apparent difference in performance of the Finkelstein algorithm between the two kinds of loci is largely an artifact of the specific realization of $(\yt_1,...,\yt_{10})$ that was used to generate this graph. The Finkelstein algorithm does not use fixed neighborhoods $\{Z_j\}$, so the distinction between central and peripheral is simply does not apply, except for the first and last loci overall.)

\begin{figure}
\includegraphics[scale=1]{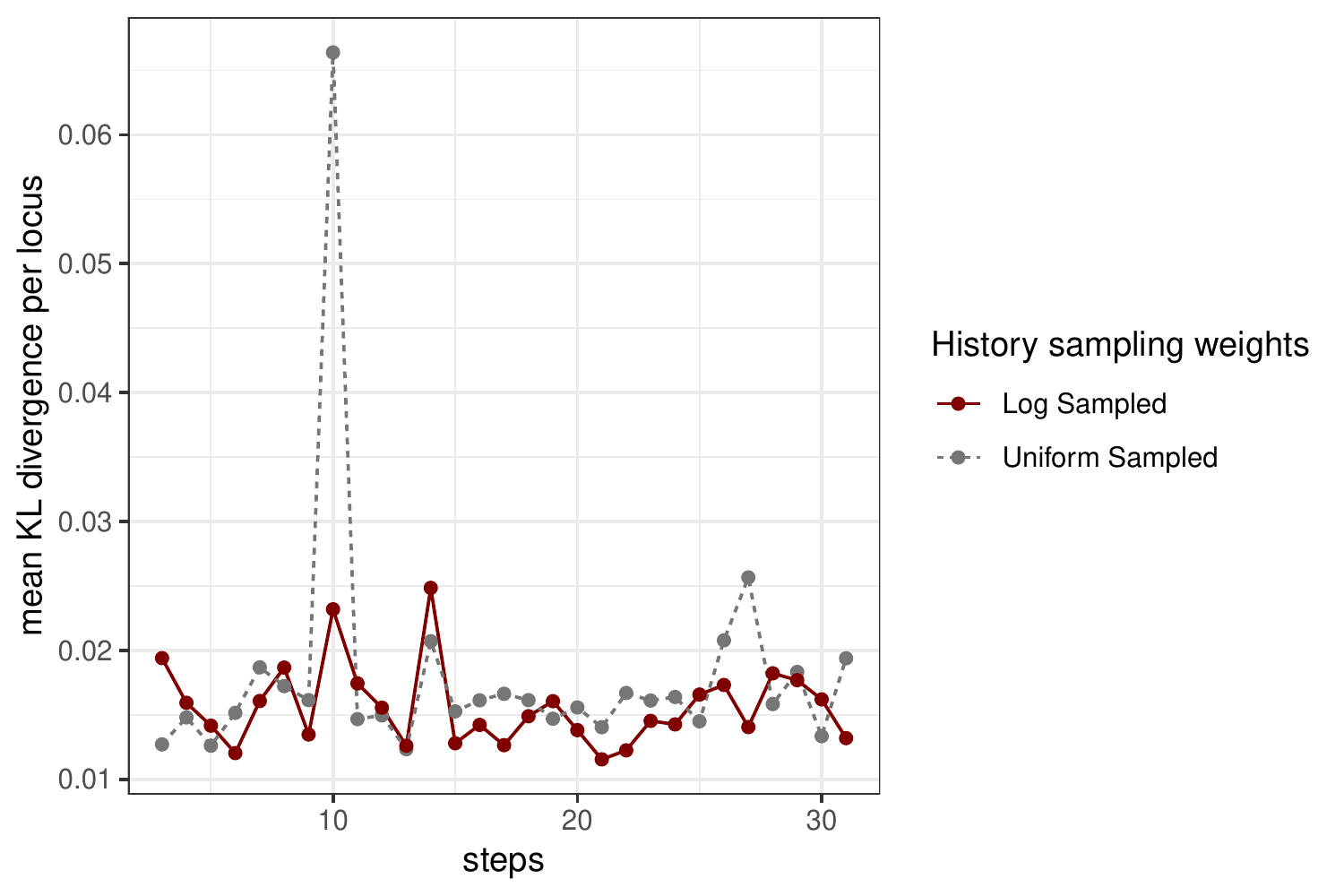}
\caption[]{Stability of KL divergence across time steps.}
\label{stabilityplot}
\end{figure}

The error of the Finkelstein algorithm, like that of the block particle filter, appears to remain stable over time, as seen in Figure~\ref{stabilityplot}. This shows the time evolution of the KL divergence between the true filtering distribution, as calculated using the Kalman filter, and a Gaussian with mean vector and covariance matrix inferred from a Finkelstein particle filter. It appears that uniform sampling ($g_\mathrm{uniform}$) can occasionally be unsuccessful in sampling good histories, as reflected by the spikes in that line; log sampling ($g_\mathrm{bentlog}$) had superior stability.

\section{Conclusion}

I have introduced the novel Finkelstein particle filtering algorithm for estimating the filtering distribution of models with high dimensionality due to large spatial extent. In such models, the simple bootstrap particle filtering algorithm is unusable. But, as with the previously-proposed block particle filter, my algorithm relies on the locality of dynamics to resolve this problem, focusing on a small area at a time.

Using simulations, I have showed that the error of means of my algorithm has lower bias but higher variance than the block particle filter, given comparable parameters. All in all, the total squared error of means of the Finkelstein algorithm is more homogeneous across loci than that of the block algorithm; lower for loci peripheral to a neighborhood in the block particle filter, but higher for those which are central. I also give empirical evidence that the error of this algorithm is stable over time, making it a candidate for online data assimilation tasks.

It is commonplace to prefer variance over bias when such a tradeoff is possible, because this allows improving precision with additional computing power by independent reruns of the algorithm. That improved precision would certainly be possible in this case with the Finkelstein algorithm. This picture is slightly complicated by the fact that such computing power might enable better results from the block particle filter by increasing the neighborhood size. But there are several problems with just increasing neighborhood size. Above all, computing power (that is, number of particles) needed could be up to exponential in neighborhood size, while it's just quadratic in number of Finkelstein particles or linear in independent Finkelstein runs. Second, unlike number of particles, neighborhood size comes in sizeable discrete intervals; it may not be possible to effectively use a small additional amount of computing power. And finally, to reduce the bias of the block particle filter, neighborhood size must be increased up-front, while the variance Finkelstein particle filter can in be reduced by independent runs (perhaps even by different scientists).   

Thus, I believe that the Finkelstein particle filter algorithm offers meaningful advantages over prior proposals. In future work, I will extend this to cover chaotic dynamics in a deterministic or quasi-deterministic model. 

\printbibliography

\end{document}